\documentclass[12pt]{article}
\begin{document}
\pagenumbering{arabic}
\begin{titlepage}
\title{Bondi-Sachs energy-momentum and the energy of gravitational radiation}
\author{J. W. Maluf$\,^{(a)}$, J. F. da Rocha-Neto$\,^{(b)}$
and S. C. Ulhoa$\,^{(c)}$}

\date{}
\maketitle
\bigskip
{\footnotesize
\centering{  Instituto de F\'{\i}sica\par
Universidade de Bras\'{\i}lia\par
C.P. 04385\par
70.919-970 Bras\'{\i}lia DF, Brazil}\par}
\bigskip
\begin{abstract}
We construct the gravitational energy-momentum of the Bondi-Sachs 
space-time, in the famework of the teleparallel equivalent of general relativity
(TEGR). The Bondi-Sachs line element describes gravitational radiation in the
asymptotic region of the space-time, and is determined by the mass aspect and by
two functions, $c$ and $d$, that yield the news functions, which are interpreted 
as the radiating degrees of freedom of the gravitational field. The standard
expression for the Bondi-Sachs energy-momentum is constructed in terms of
the mass aspect only. The expression that we obtain in the context of the
TEGR is given by the standard expression, which represents the gravitational
energy of the source, plus a new term that is determined by the two functions
$c$ and $d$. We interpret this new term as the energy of gravitational 
radiation.
\end{abstract}
\thispagestyle{empty}
\par

\bigskip
{\footnotesize
\noindent (a) wadih@unb.br, jwmaluf@gmail.com\par
\noindent (b) rocha@fis.unb.br\par
\noindent (c) sc.ulhoa@gmail.com\par}
\vfill
\end{titlepage}

\section{Introduction}

One of the most interesting consequences of general relativity is the 
description of gravitational radiation generated by isolated astrophysical 
configurations, that loose energy in the form of radiation. The space-time 
around these configurations is not strictly asymptotically flat because 
although the metric tensor components fall off in the expected way as 
$g_{\mu\nu}=\eta_{\mu\nu}+h_{\mu\nu}(1/r)$ in the asymptotic limit
$r\rightarrow \infty$, the time derivative of $h_{\mu\nu}$ is of order 
$1/r$ at spacelike infinity in Cartesian coordinates. 
The first significant work on this 
issue is due to Bondi and collaborators \cite{Bondi}, who established the
coordinates and notation that are currently employed in the analysis of
gravitational radiation. The line element obtained by Bondi and collaborators
is constructed out of the functions $M(u,\theta)$ and $c(u,\theta)$,
where $u$ is the retarded time ($u=t-r$), and $r$ and $\theta$ are spherical
coordinates. These functions do not depend on the spherical coordinate 
$\phi$ because the line element is axially symmetric. The function $M$ is 
called the mass aspect, and the time derivative 
$\partial_0 c=\partial c/\partial u$ is identified as the first news function.

Bondi's line element was subsequently generalised by Sachs \cite{Sachs}, who
abandoned the axial symmetry and obtained the most general metric tensor that 
describes gravitational radiation at spacelike and null infinities. In the work
by Sachs there also appears the function $d$, that yields the second news 
function $\partial_0 d$. The two news functions are interpreted as the 
radiating degrees of freedom of the gravitational field. The metric tensor 
obtained by Sachs yields what is presently known as the Bondi-Sachs space-time.
The mathematical expression of the Bondi-Sachs metric tensor is rather 
intricate. There are very good review articles that clarify the several 
aspects of the subject, and also explain the emergence of the Bondi-Sachs 
energy-momentum vector \cite{Trautman,Pirani,Sachs-2,Goldberg,G2,Chrusciel}.

The Bondi-Sachs energy-momentum is constructed out of the mass aspect
$M(u,\theta,\phi)$ only, i.e., it does not depend on the functions 
$c(u,\theta,\phi)$ and $d(u,\theta,\phi)$ (see Eq. (4.4) of Ref. 
\cite{Goldberg}). To some extent, it is intriguing that the total energy
and momentum do not depend on the these functions.
Recently, the total Bondi-Sachs energy-momentum has been 
compared to the ADM expression \cite{ADM} of the total gravitational 
energy-momentum at spacelike infinity of the Bondi-Sachs space-time. The 
metric tensor of the latter has been 
rewritten in the ordinary $(t,r,\theta,\phi)$ coordinates by parametrizing
the spacelike hypersurfaces by the standard time $t$. It has been found 
\cite{Z1,Z2,Z3} that the resulting expression for the total ADM energy-momentum
depends on the mass aspect, as expected, but also depends on the functions $c$
and $d$. Although the total ADM energy-momentum is 
strictly constructed for asymptotically flat space-times, the deviation of the
resulting expression from the standard Bondi-Sachs energy-momentum
- and its dependence on the functions $c$ and $d$ - is a very interesting 
result.

In this article we obtain the expression of the total gravitational 
energy-momentum of the Bondi-Sachs space-time in the context of the 
teleparallel equivalent of general relativity (TEGR) \cite{Maluf1}. The latter
is an alternative geometrical formulation of general relativity based on the 
tetrad field. The TEGR provides a natural geometrical setting for consistent
definitions of energy, momentum and angular momentum of the gravitational field.
The definitions are given by surface integrals, they satisfy conservation
equations, the algebra of the Poincar\'e group, and arise from well defined
densities \cite{Maluf2}. These definitions are possible because of the 
structure of the field equations and the covariance of the tetrad fields
under global SO(3,1) transformations. The expressions for the gravitational 
energy-momentum and angular momentum are covariant under global SO(3,1)
transformations. In special relativity the energy-momentum and angular momentum
of localised material systems are frame dependent, and so they are in the TEGR, 
since the presence of gravitational fields (in the Newtonian approximation, for
instance) do not modify this situation.

We find that the total energy-momentum of the Bondi-Sachs space-time is given
by the standard integral of the mass aspect, plus a new term that contains the
time derivative of the functions $c$ and $d$. This new term is interpreted as 
the energy of gravitational radiation. Attempts have been made in the past to 
arrive at such quantity, but failed because there was no guarantee that the 
integrals would be convergent \cite{Trautman}. The expression that we obtain
in the present geometrical framework is finite. We test our expression by using
a simple expression for the news function suggested long time go in the 
literature.

In section II we briefly present the geometrical framework of the TEGR, the
definition of the energy-momentum 4-vector, and the conservation equations.
The Bondi-Sachs line element is presented in section III. In this section we
display the asymptotic expansion of the metric tensor components and also
construct the set of tetrad fields that will be used in the calculations.
Section IV contains all relevant expressions and steps that yield the 
gravitational energy-momentum of the Bondi-Sachs space-time. The application 
to the news function suggested by Papapetrou \cite{Papapetrou}, Halliday and 
Janis \cite{HJ} and Hobill \cite{Hobill} is given in section V. Finally in 
section VI we present the final remarks.
\bigskip

\noindent {\bf Notation}:
space-time indices $\mu, \nu, ...$ and SO(3,1) (Lorentz) indices
$a, b, ...$ run from 0 to 3. Time and space indices are indicated according to
$\mu=0,i,\;\;a=(0),(i)$. The tetrad fields are represented by $e^a\,_\mu$, and 
the torsion tensor by 
$T_{a\mu\nu}=\partial_\mu e_{a\nu}-\partial_\nu e_{a\mu}$. The flat space-time 
metric tensor raises and lowers tetrad indices, and is fixed by 
$\eta_{ab}= e_{a\mu} e_{b\nu}g^{\mu\nu}=(-1,+1,+1,+1)$.
The frame components are given by the inverse tetrads 
$\lbrace e_a\,^\mu \rbrace$. The determinant of the tetrad field is written as
$e=\det(e^a\,_\mu)$.

The torsion tensor $T_{a\mu\nu}$ is sometimes related to the object of
anholonomity $\Omega^\lambda\,_{\mu\nu}$ via 
$\Omega^\lambda\,_{\mu\nu}= e_a\,^\lambda T^a\,_{\mu\nu}$.
It is important to note that we assume that the space-time geometry is defined 
by the tetrad fields only, and thus the only
possible non-trivial definition for the torsion tensor is given by
$T^a\,_{\mu\nu}$. This tensor is related to the 
antisymmetric part of the Weitzenb\"ock  connection 
$\Gamma^\lambda_{\mu\nu}=e^{a\lambda}\partial_\mu e_{a\nu}$, which
determines the Weitzenb\"ock space-time and the distant parallelism of vector
fields.

\section{A brief review of the TEGR}

The teleparallel equivalent of general relativity is a theory for the 
gravitational field based on the tetrad field. The dynamics of the gravitational
field in the TEGR is exactly the same as in the standard metric formulation of 
general relativity. The tetrad fields have 16 independent components, and the 
extra six components, compared to the ten components of the metric tensor,
allow the establishment of additional geometric structures. In the geometrical
framework determined by the tetrad fields, one may dispose of the concepts of 
both the Riemannian and Weitzenb\"{o}ck geometries. 
The equivalence of the TEGR with Einstein's general 
relativity is established by means of an identity between the scalar 
curvature $R(e)$, constructed out of the tetrad fields, and a combination of
quadratic terms of the torsion tensor, 

\begin{equation}
eR(e)\equiv -e({1\over 4}T^{abc}T_{abc}+{1\over 2}T^{abc}
T_{bac}-T^aT_a)
+2\partial_\mu(eT^\mu)\,.
\label{1}
\end{equation}
The formulation of Einstein's general relativity in the context of the
teleparallel geometry is discussed in several references, see
\cite{Maluf1,Maluf2,Hehl2,Maluf3,FG,Hehl1}. The Lagrangian density of the 
TEGR is given by the combination of the quadratic terms on the right hand side
of Eq. (1),

\begin{eqnarray}
L&=& -k e({1\over 4}T^{abc}T_{abc}+{1\over 2}T^{abc}T_{bac}-
T^aT_a) -L_M\nonumber \\
&\equiv& -ke\Sigma^{abc}T_{abc}-L_M\,, 
\label{2}
\end{eqnarray}
where $k=c^3/16\pi G$, $T_a=T^b\,_{ba}$, 
$T_{abc}=e_b\,^\mu e_c\,^\nu T_{a\mu\nu}$, and

\begin{equation}
\Sigma^{abc}={1\over 4} (T^{abc}+T^{bac}-T^{cab})
+{1\over 2}( \eta^{ac}T^b-\eta^{ab}T^c)\;.
\label{3}
\end{equation}
$L_M$ represents the Lagrangian density of the matter fields. The field 
equations derived from (2) are equivalent to Einstein's equations. They read
\cite{Maluf1}

\begin{equation}
e_{a\lambda}e_{b\mu}\partial_\nu (e\Sigma^{b\lambda \nu} )-
e (\Sigma^{b\nu}\,_aT_{b\nu\mu}-
{1\over 4}e_{a\mu}T_{bcd}\Sigma^{bcd} )={1\over {4k}}eT_{a\mu}\,,
\label{4}
\end{equation}
where
$\delta L_M / \delta e^{a\mu}=eT_{a\mu}$. 
It is possible to show that the left hand side of the equation
above may be rewritten as
${1\over 2}e\left[ R_{a\mu}(e)-{1\over 2}e_{a\mu}R(e)\right]$, which proves
the equivalence of the present formulation with the standard metric theory.

Equation (4) may be rewritten in a simplified form as

\begin{equation}
\partial_\nu(e\Sigma^{a\lambda\nu})={1\over {4k}}
e\, e^a\,_\mu( t^{\lambda \mu} + T^{\lambda \mu})\;,
\label{5}
\end{equation}
where $T^{\lambda\mu}=e_a\,^{\lambda}T^{a\mu}$, and
$t^{\lambda\mu}$ is defined by

\begin{equation}
t^{\lambda \mu}=k(4\Sigma^{bc\lambda}T_{bc}\,^\mu-
g^{\lambda \mu}\Sigma^{bcd}T_{bcd})\,.
\label{6}
\end{equation}
In view of the antisymmetry property 
$\Sigma^{a\mu\nu}=-\Sigma^{a\nu\mu}$, it follows that

\begin{equation}
\partial_\lambda
\left[e\, e^a\,_\mu( t^{\lambda \mu} + T^{\lambda \mu})\right]=0\,.
\label{7}
\end{equation}
The equation above yields the continuity (or balance) equation,

\begin{equation}
{d\over {dt}} \int_V d^3x\,e\,e^a\,_\mu (t^{0\mu} +T^{0\mu})
=-\oint_S dS_j\,
\left[e\,e^a\,_\mu (t^{j\mu} +T^{j\mu})\right]\,.
\label{8}
\end{equation}
We identify
$t^{\lambda\mu}$ as the gravitational energy-momentum tensor
\cite{Maluf3}, and

\begin{equation}
P^a=\int_V d^3x\,e\,e^a\,_\mu (t^{0\mu} 
+T^{0\mu})\,,
\label{9}
\end{equation}
as the total energy-momentum contained within a volume $V$ of the
three-dimensional space. In view of (5), Eq. (9) may be written as 

\begin{equation}
P^a=-\int_V d^3x \partial_j \Pi^{aj}\,,
\label{10}
\end{equation}
where $\Pi^{aj}=-4ke\,\Sigma^{a0j}$, which is the momentum
canonically conjugated to $e_{aj}$ \cite{Maluf1,Maluf3,Maluf4}.
The quantity $\partial_j \Pi^{aj}$ is a well defined space-time scalar
density, that transforms as a vector under the global SO(3,1) group.
The expression above may be transformed into a surface integral, and is the
definition for the gravitational energy-momentum discussed in Refs.
\cite{Maluf2,Maluf3,Maluf4}, obtained in the framework of the Hamiltonian
vacuum field equations. Note that Eq. (8) is a true energy-momentum 
conservation equation.

The emergence of a non-trivial total divergence is a feature of 
theories with torsion. The integration of this total divergence yields a
surface integral. If we consider the $a=(0)$ component of Eq. (10) in 
spherical coordinates and a spacelike surface $S$ determined by $r=$ constant,
we have

\begin{equation}
P^{(0)}={E\over c}=-\oint_S dS_j\, \Pi^{(0)j}=
4k\oint_S d\theta d\phi \,\Sigma^{(0)01}\,.
\label{11}
\end{equation}
Adopting asymptotic boundary conditions for the tetrad fields, we find 
\cite{Maluf4} that in the limit $S\rightarrow \infty$ 
the resulting expression is precisely the surface 
integral at spacelike infinity that defines the ADM energy \cite{ADM}. 
This fact is a indication that Eq. (\ref{9}) does indeed represent the 
gravitational energy-momentum vector. But note that there is no restriction
regarding the applicability of definition (\ref{10}). The latter may be applied
to arbitrary space-times, with arbitrary boundary conditions.

\section{The Bondi-Sachs space-time and the tetrad fields}

The Bondi-Sachs metric tensor describes gravitational radiation at null and
spatial infinities, and in both asymptotic limits the the metric tensor 
$g_{\mu\nu}$ approaches the flat space-time metric tensor $\eta_{\mu\nu}$, but
time derivatives of $g_{\mu\nu}$ fall off as $1/r$. As a consequence, 
gravitational waves may in principle be detected at spatial or null infinities.
The line element of the Bondi-Sachs space-time
in spherical $(u,r,\theta,\phi)$ coordinates, where $u=t-r$ is
the retarded time, is constructed out of the functions $M(u,\theta,\phi)$, 
$c(u,\theta,\phi)$ and $d(u,\theta,\phi)$. It is given by

\begin{eqnarray}
ds^2&=&g_{00}\,du^2+g_{22}\,d\theta^2+g_{33}\,d\phi^2 \nonumber \\
&{}&+2g_{01}\,du\,dr+2g_{02}\,du\,d\theta+2g_{03}\,du\,d\phi+
2g_{23}d\theta\,d\phi\,, 
\label{12}
\end{eqnarray}
where

\begin{eqnarray}
g_{00}&=&{V\over r}e^{2\beta}-r^2(e^{2\gamma}U^2\cosh 2\delta 
+e^{-2\gamma}W^2\cosh 2\delta +2UW \sinh 2\delta)\,, \nonumber \\
g_{01}&=& -e^{2\beta}\,, \nonumber \\
g_{02}&=&-r^2(e^{2\gamma}U\cosh 2\delta+W\sinh 2\delta)\,, \nonumber \\
g_{03}&=&-r^2\sin\theta(e^{-2\gamma}W\cosh 2\delta 
+U \sinh 2\delta)\,, \nonumber \\
g_{22}&=&r^2e^{2\gamma}\cosh 2\delta\,, \nonumber \\
g_{33}&=&r^2e^{-2\gamma}\cosh 2\delta\,\sin^2 \theta\,, \nonumber \\
g_{23}&=&r^2\sinh 2\delta\,\sin\theta\,.
\label{13}
\end{eqnarray}
We adopt the usual convention $(u,r,\theta,\phi)=(x^0,x^1,x^2,x^3)$.
The functions $\beta$, $\gamma$, $\delta$, $U$ and $W$ in the equations above
are given only in asymptotic form, in powers of $1/r$. 

In this article we will
present the asymptotic expansions of all field quantities that are effectively
needed in the calculations, i.e., we will dispense with the powers of $1/r$ of 
the field quantities that do not contribute to the calculations. Thus, the 
asymptotic form of the functions above are

\begin{eqnarray}
V&\simeq& -r+2M\,, \nonumber \\
\beta &\simeq& -{{c^2+d^2}\over {4r^2}}\,,\nonumber \\
\gamma &\simeq& {c\over r}\,, \nonumber \\
\delta &\simeq& {d\over r}\,, \nonumber \\
U&\simeq& -{{l(u,\theta,\phi)}\over r^2}\,, \nonumber \\
W&\simeq& -{{\bar{l}(u,\theta,\phi)} \over r^2}\,,
\label{14}
\end{eqnarray}
where 
$$l=\partial_2 c+2c\,\cot\theta+\partial_3 d\,\csc \theta\,,$$
$$\bar{l}=\partial_2 d +2d\,\cot\theta -\partial_3 c\csc \theta\,.$$
In the limit $r\rightarrow \infty$, the asymptotic form of the functions above
yield

\begin{eqnarray}
g_{00}&\simeq & -1+{{2M}\over r}\,, \nonumber \\
g_{01}&\simeq & -1+{{c^2+d^2}\over {2r^2}}\,,\nonumber \\
g_{02}&\simeq & l+{1\over r}(2cl +2d\bar{l} -p)\,, \nonumber \\
g_{03}&\simeq & \bar{l}\sin\theta + 
{1\over r}(-2c\bar{l}+2dl -\bar{p})\sin\theta\,, \nonumber \\
g_{22}&\simeq & r^2+ 2cr+2(c^2+d^2)\,, \nonumber \\
g_{33}&\simeq & \lbrack r^2- 2cr+2(c^2+d^2)\rbrack\sin^2\theta\,, \nonumber \\
g_{23}&\simeq & 2dr\sin\theta+{{4d^3}\over {3r}} \sin\theta\,.
\label{15}
\end{eqnarray}
The functions $p$ and $\bar{p}$ are defined in Refs. \cite{Z1,Z2}. They depend
on functions that are not defined above, but since they will not contribute to 
the final expressions, we will not present their definitions here.

The expressions of the contravariant components of the metric tensor are
calculated by means of the standard procedure out of Eqs. (\ref{13}) (not
out of Eqs. (\ref{15})). The inverse components are given by 
$g^{\mu\nu}=(-1)^{\mu+\nu}(1/g)\,M_{\mu\nu}$,  where 
$g=-g_{01}^2(g_{22}g_{33}-g_{23}^2)$ is the determinant of the metric tensor,
and $M_{\mu\nu}$ is the co-factor of the $\mu\nu$ component (we have taken into
account all necessary powers of $1/r$ of the functions given in Eq. (\ref{14})).
We find

\begin{eqnarray}
g^{00}&=& g^{02}=g^{03}=0\,,\nonumber \\
g^{01}&\simeq & -1-{{c^2+d^2}\over {2r^2}}\,,\nonumber \\
g^{11}&\simeq &1-{{2M}\over r}\,,\nonumber \\
g^{12}&\simeq &{l\over r^2}\,,\nonumber \\
g^{13}&\simeq &{{\bar{l}\sin\theta}\over r^2}\,,\nonumber \\
g^{22}&\simeq &{1\over r^2}\,,\nonumber \\
g^{33}&\simeq &{1\over {r^2\sin^2\theta}}\,,\nonumber \\
g^{23}&\simeq & -{{2d}\over {r^3\sin^2\theta}}\,.
\label{16}
\end{eqnarray}

Now we turn to the construction of the tetrad fields. The inverse tetrads
$e_a\,^\mu$ determine the frame adapted to a particular class of observers in
space-time. Let the curve $x^\mu(\tau)$ represent the timelike worldline $C$ 
of an observer in space-time, where $\tau$ is the proper time of the observer.
The velocity of the observer along $C$ is given by $u^\mu=dx^\mu/d\tau$.
A frame adapted to this observer is constructed by identifying the
timelike component of the frame $e_{(0)}\,^\mu$ with the velocity $u^\mu$
of the observer: $e_{(0)}\,^\mu=u^\mu(\tau)$. The three other 
components of the frame, $e_{(i)}\,^\mu$, are orthogonal to $e_{(0)}\,^\mu$, and
may be oriented in the
three-dimensional space according to the symmetry of the physical system.
A static observer in space-time is defined by the condition $u^\mu=(u^0,0,0,0)$.
Thus, a frame adapted to a static observer in space-time must satisfy the
conditions $e_{(0)}\,^i(t,x^k)=(0,0,0)$. It is easy to verify, by means of a
coordinate transformation, that in terms of the retarded time $u$ we also have
$e_{(0)}\,^i(u,x^k)=(0,0,0)$. The Bondi-Sachs space-time is not axially
symmetric, and therefore there are no distinguished directions at spacelike 
infinity. Since we will evaluate surface integrals at spacelike infinity, i.e.,
we will be interested only in total quantities (we will also integrate over a 
surface $S$ determined by $r=$ constant, for $r$ finite but sufficiently large),
any set of tetrad fields that satisfy the asymptotic expansion 
$e_{a\mu}\simeq \eta_{a\mu}+(1/2)h_{a\mu}(1/r)$
in Cartesian coordinates when $r\rightarrow \infty$, and that satisfy the 
conditions $e_{(0)}\,^i(u,r,\theta,\phi)=0$, will serve our purposes.
For such a frame, $e_{(1)}\,^\mu$, $e_{(2)}\,^\mu$ and $e_{(3)}\,^\mu$ will 
define the usual unit frame vectors in the $x$, $y$ and $z$ directions, 
respectively, in the limit $r\rightarrow \infty$, provided $e_a\,^\mu$ is 
constructed in Cartesian coordinates. If we restrict the Bondi-Sachs metric to 
the Bondi metric tensor (by making $d=0=\bar{l}$), then the latter is axially 
symmetric and $e_{(3)}\,^\mu$ will define
the unit vector in the $z$ direction at spacelike infinity.

It is not straightforward to construct a simple set of tetrad fields that yields
Eq. (\ref{12}), and that satisfy the conditions $e_{(0)}\,^i=0$. Note that 
$e_{(0)}\,^i=0$ implies $e^{(i)}\,_0=0$. One such set of tetrad fields that 
satisfy these requirements, and acquires the asymptotic form 
$e_{a\mu}\simeq\eta_{a\mu}+(1/2)h_{a\mu}$ at spacelike infinity, is given in
$(u,r,\theta,\phi)$ coordinates by

\begin{eqnarray}
e_{(0)\mu}&=&(-A,-E,-F,-G)\,, \nonumber \\
e_{(1)\mu}&=& (0, \,
B_1\sin\theta \cos\phi+B_2\cos\theta \cos\phi-B_3\sin\theta \sin\phi\,, 
\nonumber \\
& & C_1 r \cos\theta \cos\phi-C_2 \sin\theta \sin\phi\,, \nonumber \\
& & -Dr\sin\theta \sin\phi)\,,\nonumber \\
e_{(2)\mu}&=& (0, \,
B_1\sin\theta \sin\phi+B_2\cos\theta \sin\phi+B_3\sin\theta \cos\phi\,, 
\nonumber \\
& & C_1 r \cos\theta \sin\phi-C_2 \sin\theta \cos\phi\,, \nonumber \\
& & Dr\sin\theta \cos\phi)\,,\nonumber \\
e_{(3)\mu}&=&(0,\,B_1\cos\theta-B_2\sin\theta, -C_1r\cos\theta,0)\,.
\label{17}
\end{eqnarray}

The quantities $A,B_1,B_2,B_3,C_1,C_2,D,E,F,G$ are determined by requiring that
$e_{a\mu}$ yields the metric tensor components (\ref{15}) according to
$e_{a\mu}e_{b\nu}\eta^{ab}=g_{\mu\nu}$. The determination of
exact form of these quantities in terms of the metric tensor components 
(\ref{13}) is very complicated and useless for our purposes. We will need the
components of the torsion tensor only in the asymptotic limit 
$r \rightarrow \infty$. These quantities must satisfy the following equations,

\begin{eqnarray}
-A^2&=& g_{00}\nonumber \\
-AE&=&g_{01} \nonumber \\
-AF&=&g_{02} \nonumber \\
-AG&=&g_{03} \nonumber \\
-E^2+B_1^2+B_2^2+B_3^2\sin^2\theta&=&g_{11}=0\nonumber \\
-EF+B_2(C_1r)+B_3C_2\sin^2\theta&=&g_{12}=0\nonumber \\
-EG+B_3(Dr)\sin^2\theta&=&g_{13}=0 \nonumber \\
-F^2+(C_1r)^2+C_2^2\sin^2\theta&=&g_{22} \nonumber \\
-G^2+(Dr)^2\sin^2\theta&=&g_{33} \nonumber \\
-FG+C_2(Dr)\sin^2\theta&=& g_{23}\,.
\label{(18)}
\end{eqnarray}

As we mentioned earlier, in the asymptotic expansion of the field quantities we
will display the terms only up to the power of $1/r$ that is actually needed 
in the calculations, taking care that we do not neglect any relevant term up
to $(1/r)^2$. Terms of order $(1/r)^n$, with $n\geq 3$, do not contribute to the
final, total expressions. We find

\begin{eqnarray}
A&\simeq& 1-{M\over r}\,, \nonumber \\
E&\simeq& 1+{M\over r}\,,\nonumber \\ 
F&\simeq& -l-{1\over r}(2cl+2d\bar{l}+Ml-p)\,, \nonumber \\
G&\simeq& -\sin\theta \biggl[ \bar{l}+
{1\over r}(-2c\bar{l}+2dl+M\bar{l}-\bar{p})\biggr]\,, \nonumber \\
B_1 &\simeq& 1+{M\over r}\,, \nonumber \\
B_2 &\simeq& -{l\over r}-{1\over r^2}(2Ml+cl-p)\,, \nonumber \\
B_3&\simeq& -{1\over {\sin\theta}}\biggl[ {{\bar{l}\over r}}+{1\over r^2}
(2M\bar{l}-c\bar{l}+2dl-\bar{p})\biggr]\,,\nonumber \\
C_1&\simeq& 1+{c\over r}+{1\over r^2}
\biggl[{{l^2}\over 2}+c^2-d^2\biggr]\,,\nonumber \\
C_2&\simeq& {1\over {\sin\theta}}\biggl[ 2d+{1\over r}(l\bar{l}+2cd)\biggr]\,.
\nonumber \\
D&\simeq& 1-{c\over r}+{1\over r^2}\biggl( {{\bar{l}^2\over 2}}+
c^2+d^2\biggr)\,.
\label{19}
\end{eqnarray}
The expressions above completely fix the set of tetrad fields given by 
Eq. (\ref{17}). It must be noted that the $\sin\theta$ in the denominator of 
$B_3$ and $C_2$ in the expressions above does not imply a divergence of 
$e_{a\mu}$ on the $z$ axis of the coordinate system, when $\theta =0$. In Eq. 
(\ref{17}) these quantities are multiplied by $\sin\theta$. It can be shown 
that in $(u,x,y,z)$ or in $(t,x,y,z)$ coordinates, the set of tetrad fields 
given by Eq. (\ref{17}) is everywhere smooth in the three-dimensional space.

\section{The gravitational energy-momentum}

\subsection{Gravitational energy}

The gravitational energy contained within a two-dimensional spacelike surface
$S$ (defined by $r=$ constant) in an arbitrary space-time, in spherical 
coordinates, is obtained by evaluating the quantity $\Sigma^{(0)01}$, as 
indicated in Eq. (\ref{11}). The simplification of this quantity is 
crucial to arrive at the final result. Taking into account Eq. (\ref{3}), and 
the fact that $g^{00}=g^{02}=g^{03}=0$, we find

\begin{eqnarray}
\Sigma^{(0)01}&=&e^{(0)}\,_\mu \Sigma^{\mu 01}\nonumber \\
&=&e^{(0)}\,_0\Sigma^{001}+e^{(0)}\,_1\Sigma^{101}+e^{(0)}\,_2\Sigma^{201}+
e^{(0)}\,_3\Sigma^{301}\nonumber \\
&=&A\biggl[{1\over 4}\biggl(T^{001}+T^{001}-T^{100}\biggr)+
{1\over 2}g^{01}T^0 \biggr]\nonumber \\
&+ &E\biggl[{1\over 4}\biggl(T^{101}+T^{011}-T^{110}\biggr)+
{1\over 2}\biggl(g^{11}T^0-g^{01}T^1\biggr)\biggr] \nonumber \\
&+&F\biggl[{1\over 4}\biggl(T^{201}+T^{021}-T^{120}\biggr)+
{1\over 2}g^{21}T^0\biggr] \nonumber \\
&+&G\biggl[{1\over 4}\biggl(T^{301}+T^{031}-T^{130}\biggr)+
{1\over 2}g^{31}T^0 \biggr]\,.
\label{20}
\end{eqnarray}
where $T^\mu=T^\alpha\,_\alpha\,^\mu$. 

In Eq. (\ref{11}), $\Sigma^{(0)01}$ is multiplied by the determinant $e$,
whose asymptotic expression is $e\simeq r^2\sin\theta$. In order to obtain
a non-vanishing value for the gravitational energy when the surface $S$ of
integration approaches the limit $r\rightarrow \infty$, we must select
the terms of $\Sigma^{(0)01}$ that are of the order $1/r^2$. Terms of the
order $1/r^n$, with $n \ge 3$, do not contribute to the final expression. 

By expanding Eq. (\ref{20}) in terms of the covariant torsion tensor 
$T_{\alpha\mu\nu}$, we observe that all terms will be of the type
$g^{\mu\alpha}g^{0\beta}g^{1\gamma}T_{\alpha\beta\gamma}$. In view of 
Eq. (\ref{16}),  we see that all products of the type
$g^{\mu\alpha}g^{0\beta}g^{1\gamma} $ that arise in the expansion
of Eq. (\ref{20}) are at least of the order 
$1 /r^2$. Products of the order $1/ r^n$, with $n\ge 3$, yield a
vanishing contribution in the limit $r\rightarrow \infty$. Therefore we 
keep only the products that are of the order $1 /r^2$, and consider the 
contributions from 
$T_{\alpha\mu\nu}$ whose values at spacelike infinity are of the order
$1/r^0$. It is important to mention that there does not arise any divergent
term (of order $O(r)$ or higher) in the expansion of $e\Sigma^{(0)01}$.
The final expression turns out to be finite. After long calculations, we
conclude that the non-vanishing value of $\Sigma^{(0)01}$ are simplified to

\begin{eqnarray}
\Sigma^{(0)01}&=& A\,\Sigma^{001}+E\,\Sigma^{101}+F\,\Sigma^{201}+
G\,\Sigma^{301}\nonumber \\
&=&
-{1\over 2} A\,(g^{01}g^{01}g^{22}T_{212}+g^{01}g^{01}g^{33}T_{313})\nonumber\\
&&+ {1\over 2} E\, (g^{01}g^{01}g^{22}T_{202}+g^{01}g^{01}g^{33}T_{303}) 
\nonumber \\
&&+{1\over 4} F\, g^{01}g^{01}g^{22}(T_{012}-T_{201}-T_{102})\nonumber \\
&&+{1\over 4} G\, g^{01}g^{01}g^{33}(T_{013}-T_{301}-T_{103})\,.
\label{21}
\end{eqnarray}

Inspection of the right hand side of the equation above indicates that we need
to calculate only 10 components of $T_{\alpha\mu\nu}$. Dispensing with terms 
of order $(1/r)^2$ and higher orders, these components are given asymptotically
by

\begin{eqnarray}
T_{201}&\simeq& -\partial_0 l +{1\over r}\biggl[ -c\partial_0 l -\partial_0
(cl+2Ml)-2d\partial_0 \bar{l}+\partial_0 p\biggr]\,, \nonumber \\
T_{301}&\simeq& -(\partial_0 \bar{l}) \sin\theta+{{\sin\theta}\over r}\biggl[
\bar{l}\partial_0 M +c\partial_0 \bar{l}-\partial_0(2M\bar{l}-c\bar{l}
+2dl)+\partial_0 \bar{p}\biggr]\,, \nonumber \\
T_{102}&\simeq& \partial_0 l+{1\over r}\biggl[ -\partial_2 M +M\partial_0 l+
\partial_0(2cl +2d\bar{l}+Ml)-l\partial_0 c-2\bar{l}\partial_0 d\biggr]
\,,\nonumber \\
T_{202}&\simeq& -l\partial_0 l+r\partial_0 c+ c\partial_0 c+4d\partial_0d\,,
\nonumber \\
T_{103}&\simeq&(\partial_0) \sin\theta\, \nonumber \\
&&+{{\sin\theta}\over r}\biggl[
-\partial_3 M +\bar{l}\partial_0 c+M\partial_0 \bar{l}+
\partial_0(M\bar{l}-2c\bar{l}+2dl)-\partial_0 \bar{p}\biggr]\,, \nonumber \\
T_{303}&\simeq&-(\bar{l}\partial_0\bar{l}+ r\partial_0 c-
c\partial_0 c)\sin^2\theta\,, \nonumber \\
T_{012}&\simeq& {{\partial_2 M}\over r}\,,\nonumber \\
T_{212}&\simeq&\partial_2 l -M\,,\nonumber \\
T_{013}&\simeq& {{\partial_3 M}\over r}\,, \nonumber \\
T_{313}&\simeq& (\partial_3 \bar{l})\sin\theta-M\sin^2\theta+
l\sin\theta\cos\theta\,. 
\label{22}
\end{eqnarray}

No term of order $1/r$ in the expressions above will contribute to $P^{(0)}$.
However, they will contribute to the momenta $P^{(i)}$.
The substitution of Eqs. (\ref{16}) and (\ref{22}) into (\ref{21}) yields

\begin{eqnarray}
\Sigma^{(0)01}&=&{1\over r^2}\biggl[ M-{1\over 2}\biggl(
\partial_2 l+{{\partial_3\bar{l}}\over {\sin\theta}}+
l{{\cos\theta}\over {\sin\theta}}\biggl)\nonumber \\
&& -{1\over 2}l\partial_0 l -{1\over 2}\bar{l}\partial_0\bar{l}
+c\partial_0c +2d\partial_0 d\biggr]\,.
\label{23}
\end{eqnarray}
Assuming $\bar{l}(\phi)=\bar{l}(\phi+2\pi)$ and $l\sin\theta=0$ for $\theta=0$ 
and $\theta=\pi$, it is easy to see that, under 
integration, the second term in the expression above vanishes,

\begin{equation}
\int_0^{2\pi}\int_0^{\pi}d\theta d\phi \sin\theta \biggl(
\partial_2 l+{{\partial_3\bar{l}}\over {\sin\theta}}+
l{{\cos\theta}\over {\sin\theta}}\biggl)=0\,.
\label{24}
\end{equation}

Finally, substitution of Eq. (\ref{23}) and $e=r^2\sin\theta$ into definition
(\ref{11}), in the limit $r\rightarrow \infty$,  results in

\begin{equation}
P^{(0)}=4k\int_0^{2\pi} d\phi \int_0^{\pi}d\theta \sin\theta
\biggl[M+\partial_0 F\biggr]\,, 
\label{25}
\end{equation}
where

\begin{equation}
F=-{1\over 4}\biggl(l^2 + \bar{l}^2 \biggr) +{1\over 2}c^2 +d^2\,.
\label{26}
\end{equation}
The integral of $M(u,\theta,\phi)$ yields the mass aspect, but the total value
of the gravitational energy depends also on the functions $c(u,\theta,\phi)$ and
$d(u,\theta,\phi)$, as one would a priori expect. The new term $\partial_0 F$
generalises the standard Bondi-Sachs energy.

For finite values of the ordinary time $t$, the limit $r\rightarrow \infty$ 
corresponds to $u\rightarrow -\infty$, and therefore in Eq. (\ref{25}) 
we have $P^{(0)}(u=-\infty)$. However, this is not the general case. We may
consider large but finite values of the radial coordinate $r$ 
($r\gg M$ and $r\gg \partial_0F$) such that 
Eq. (\ref{23}) is verified. In this case, $P^{(0)}$ depends on arbitrary and
finite values of $u$.

In equations (\ref{22}), we observe that $T_{202}$ and $T_{303}$ depend
linearly on the radial coordinate $r$. This dependence could, in principle, 
lead to a divergent expression for $P^{(0)}$. However, the two 
contributions cancel each other, and the final expression turns out to be
finite.

In order to check the consistency of the result above, we calculated 
$P^{(0)}$ out of a slightly different set of tetrad fields that satisfies the 
same requirements that led to Eq. (\ref{17}), namely, that the frame is adapted
to a static observer and that the set of tetrad fields has the asymptotic form
$e_{a\mu}\simeq\eta_{a\mu}+(1/2)h_{a\mu}$ in the limit $r\rightarrow \infty$.
We considered $e_{a\mu}$ given by

\begin{eqnarray}
e_{(0)\mu}&=&(-A,-E,-F,-G)\,, \nonumber \\
e_{(1)\mu}&=& (0, \,
B_1\sin\theta \cos\phi+B_2\cos\theta \cos\phi-B_3\sin\theta \sin\phi\,, 
\nonumber \\
& & C r \cos\theta \cos\phi\,, \nonumber \\
& & -D_1 r\sin\theta \sin\phi +D_2 \cos\theta\cos\phi)\,,\nonumber \\
e_{(2)\mu}&=& (0, \,
B_1\sin\theta \sin\phi+B_2\cos\theta \sin\phi+B_3\sin\theta \cos\phi\,, 
\nonumber \\
& & C r \cos\theta \sin\phi\,, \nonumber \\
& & D_1 r\sin\theta \cos\phi +D_2\cos\theta\sin\phi)\,,\nonumber \\
e_{(3)\mu}&=&(0,\,B_1\cos\theta-B_2\sin\theta,-C_1r\cos\theta,-D_2\sin\theta)\,.
\label{27}
\end{eqnarray}
The values of $A,B,\cdot\cdot\cdot$ in the equation above are of course 
different from Eq. (\ref{19}), but eventually we obtain $P^{(0)}$ exactly as 
given by Eq. (\ref{25}). The latter is, therefore, the gravitational energy that
a static observer would measure at large distances from the source determined
by $M(u,\theta,\phi)$. Assuming the speed of light $c=1$ as well as $G=1$, the 
quantity 

\begin{equation}
E_{rad}={1\over {4\pi}}\int_0^{2\pi} d\phi \int_0^{\pi}d\theta \sin\theta\,
(\partial_0 F)\,,
\label{28}
\end{equation}
may be interpreted as the energy of the gravitational radiation.

\subsection{Gravitational momenta}

The evaluation of the gravitational momenta $P^{(i)}$ requires the calculation
of $\Sigma^{(i)01}=e^{(i)}\,_\mu \Sigma^{\mu 01}$. The expressions of 
$\Sigma^{\mu 01}$ were already obtained in the evaluation of $P^{(0)}$. These
are the quantities that arise on the right hand side of Eq. (\ref{21}). In the
course of the calculations we find that the radial dependence of the tetrad
components $e_{(i)2}$ and $e_{(i)3}$ in Eq. (\ref{17}) requires
the values of some components of $T_{\alpha\mu\nu}$ of order $1/r$, which are
already presented in Eq. (\ref{22}). 
As we mentioned earlier, the contributions from the functions $p$ and $\bar{p}$
cancel out in the calculations. After a number of simplifications and 
cancellations, we arrive at

\begin{eqnarray}
P^{(1)}&=& 4k\int_0^{2\pi}d\phi\int_0^\pi d\theta \biggl[
\sin^2\theta \cos\phi(\partial_0 F)\nonumber \\
&& +{1\over 4}\biggl(\sin\theta \cos\theta \cos\phi(2\partial_2 M+l\partial_0 M)
\nonumber \\
&&-\sin\phi(2\partial_3 M+\bar{l} \sin\theta \partial_0 M)\biggl)\biggr]\,,
\nonumber \\
P^{(2)}&=& 4k\int_0^{2\pi}d\phi\int_0^\pi d\theta \biggl[
\sin^2\theta \sin\phi(\partial_0 F)\nonumber \\
&& +{1\over 4}\biggl(\sin\theta \cos\theta \sin\phi(2\partial_2 M+l\partial_0 M)
\nonumber \\
&&+\cos\phi(2\partial_3 M+\bar{l} \sin\theta \partial_0 M)\biggl)\biggr]\,,
\nonumber \\
P^{(3)}&=&4k\int_0^{2\pi}d\phi\int_0^\pi d\theta \biggl[
\sin\theta \cos\theta(\partial_0 F)\nonumber \\
&& -{1\over 4} \sin^2\theta(2\partial_2 M+l\partial_0 M)\biggr]\,.
\label{29}
\end{eqnarray}
The expressions above may be simplified by making integrations by parts in the
angular variables. The expressions are further simplified by introducing the
three-dimensional vectors,

\begin{eqnarray}
\hat{r}^i&=&(\sin\theta\cos\phi,\, \sin\theta\sin\phi,\,\cos\theta)\,, 
\nonumber \\
\hat{\theta}^i&=&(\cos\theta\cos\phi,\, \cos\theta\sin\phi,\, -\sin\theta)\,, 
\nonumber \\
\hat{\phi}^i&=&(-\sin\phi,\, \cos\phi,\,0)\,.
\label{30}
\end{eqnarray}
We finally obtain

\begin{eqnarray}
P^{(i)}&=& 4k\int_0^{2\pi}d\phi \int_0^{\pi}d\theta \sin\theta \biggl[
(M+\partial_0 F) \hat{r}^i \nonumber \\
&& +{1\over 4} (l\partial_0 M)\hat{\theta}^i +
{1\over 4}(\bar{l}\partial_0 M)\hat{\phi}^i \biggr]\,.
\label{31}
\end{eqnarray}

The expression above and $P^{(0)}$ given by Eq. (\ref{25}) constitute
the gravitational energy-momentum of the Bondi-Sachs space-time in the realm of
the TEGR. One important observation is the following. It is well known that the
time derivative $\partial_0 M$ may be written, in view of the field equations,
in terms of the time derivatives of the functions $c$ and $d$ 
\cite{Sachs,Sachs-2} according to

\begin{equation}
\partial_0 M=-\lbrack (\partial_0c)^2+(\partial_0 d)^2\rbrack
+{1\over 2}\partial_0\biggl(\partial_2 l+l\cot\theta+
{{\partial_3 \bar{l}}\over {\sin\theta}}\biggr)\,.
\label{32}
\end{equation}
If we evaluate all integrals in $P^a$
in the limit $r\rightarrow \infty$, we are actually
taking the limit $u\rightarrow -\infty$. If we further assume that the news
functions satisfy the initial conditions

\begin{equation}
\partial_0 c(-\infty)=0\,, \;\;\;\;\;\; \partial_0d(-\infty)=0\,,
\label{33}
\end{equation}
then the total energy-momentum $P^a=(P^{(0)},P^{(i)})$ given by (\ref{25}) and
(\ref{31}) reduces to the well known expression for the Bondi-Sachs 
energy-momentum. These initial conditions might be physically reasonable, but
there is no justification for assuming them. We asserted above that we can 
evaluate $(P^{(0)},P^{(i)})$ over a surface $S$ of integration sufficiently 
large from the source, such that Eq. (\ref{23}) is verified. This fact allows 
us to dispense with the initial conditions given by Eq. (\ref{33}).

\section{News function in an axially symmetric space-time}

In this section we will discuss an application of Eq. (\ref{28}) by
requiring the space-time geometry to be axially symmetric.  The restriction to
such symmetry is achieved by enforcing $d=0=\bar{l}$. The latter equations imply
that the components of the space-time metric tensor do not depend on the 
variable $\phi$, ensuring the axial symmetry of the configuration. This 
simplification allows the use a general form of the news function suggested by
Papapetrou \cite{Papapetrou}, Halliday and Janis \cite{HJ} and Hobill 
\cite{Hobill}. It is very difficult to obtain the exact form of the news 
function $c(u,\theta)$ because the latter must be consistent with the mass 
coefficient $M(u,\theta)$ such that Eq. (\ref{32}) is verified, and the 
verification of the latter equation is non-trivial. 

Adopting the usual convention $x\equiv \cos\theta$, the suggested form of the
news function is written as \cite{Hobill}

\begin{equation}
c(u,x)=(1-x^2)\sum_{n=2}a_n(u){{d^2}\over {d x^2}}P_n(x)\,,
\label{34}
\end{equation}
where $P_n(x)$ are the Legendre polynomials and $a_n(u)$ are general functions
of the retarded time $u$, that reduce to constants when the source is not 
radiating. Of course, it is assumed that the summation in the expression above
(as well as all summations in this section) converges.
Hobill \cite{Hobill} specified a form of the time dependence of 
$a(u)$ and proceeded to obtain a particular expression for $c(u,x)$. Below we 
will consider the more general form given by Eq. (\ref{34}).

The function $F$ given by Eq. (\ref{26}) may be rewritten as

\begin{equation}
F=-\frac{(\partial_\theta c)^2}{4}-\cot\theta\, c(\partial_\theta c)-\frac{3}{2}\,c^2\cot^2\theta+\frac{c^2}{2\sin^2\theta}\,.
\label{35}
\end{equation}
With the help of identities involving the Legendre polynomials, we write
$c(u,x)$ in the form

\begin{equation}
c(u,\theta)=\sum_{n=2} a_n(u)
\biggl[2x{d\over {{dx}}}P_n(x)-n(n+1)P_n(x)\biggr]\,.
\label{36} 
\end{equation}
Substitution of Eq. (\ref{35}) into definition (\ref{28}) yields

\begin{equation}
E_{rad}=\frac{1}{8}\partial_0\int_{-1}^{1}\left[-(1-x^2)(\partial_x c)^2+
4x\,c\,\partial_x c+2c^2(1-3x^2)\right]dx\,.
\label{37}
\end{equation}
The equation above may be further simplified by taking into account Eq. 
(\ref{36}). We arrive at

\begin{equation}
E_{rad}=\frac{1}{8}\partial_0\left[-\int_{-1}^{1}(1-x^2)(\partial_x c)^2dx-6\sum_{n,m}a_na_m\int_{-1}^{1}x^2P_n^2P_m^2dx\right]\,,
\label{38}
\end{equation}
where $P^m_n$ are the associated Legendre polynomials. Finally, using

$$\partial_x c=\sum_{n=2} a_n\left[\frac{P_n^3}{(1-x^2)^{1/2}}-
\frac{2xP_n^2}{(1-x^2)}\right]\,,$$
we obtain

\begin{eqnarray}
E_{rad}&=&-\frac{1}{8}\partial_0\sum_n\frac{a_n^2(n+2)!
(16n^4+28n^3+20n^2+11n-105)}{(n-2)!(2n+1)(2n-1)(2n+3)}\nonumber \\
&&+\frac{1}{8}\partial_0\sum_{n,m}\frac{a_na_m(n+2)![1+(-1)^{n+m}]}{(n-2)!}
\nonumber\\
&&-\frac{3}{2}\partial_0\sum_n\frac{a_na_{n+2}(n+4)!}
{(n-2)!(2n+1)(2n+3)(2n+5)}\,.
\label{39}
\end{eqnarray}

The Bondi mass $m(u)$ is defined as the usual integral over the mass aspect
$M(u,\theta)$,

\begin{equation}
m(u)={1\over 2} \int_{-1}^{1}dx\,M(u,x)\,.
\label{40}
\end{equation}
It has been demonstrated \cite{Papapetrou,HJ,Hobill} that if one requires the
total variation of the Bondi mass, due to gravitational radiation, to be equal 
to the total variation of the mass aspect, then the summation in Eqs. 
(\ref{38}) and (\ref{39}) cannot have a finite number of terms, i.e., the 
summation cannot be truncated. 

The most interesting feature of Eq. (\ref{39}) is the 
dependence of $E_{rad}$ on a quadratic combination of the coefficients $a_n(u)$.
It might be possible to impose conditions on the coefficients $a_n(u)$ such that
the summation simplifies considerably, but we have not attempted to carry out
any simplification, which would require an analysis of the field equations.

\section{Conclusions}

The Bondi-Sachs space-time is determined by three functions, $M$, $c$ and $d$,
which depend on $(u,\theta,\phi)$. The standard expression of the 
gravitational energy of the Bondi-Sachs space-time is the integral of the
mass aspect 
$M(u,\theta,\phi)$ in the angular variables, and restricting to the Bondi 
space-time, it is identified with $m(u)$ given by Eq. (\ref{40}). It is not
clear why both expressions do not depend on the functions $c$ and $d$. 
The expression that we obtained in the realm of the TEGR, Eq. (\ref{25}), does
depend on $M$, $c$ and $d$. 
We also obtained contributions of the functions  $c$ and $d$ to the total
gravitational momenta, as given by Eq. (\ref{31}). Altogether, Eqs. (\ref{25})
and (\ref{31}) constitute the gravitational energy-momentum of the Bondi-Sachs
space-time in the framework of the TEGR.

The total energy-momentum of the Bondi-Sachs 
space-time, evaluated by means of the ADM definitions in $(t,r,\theta,\phi)$
coordinates, does depend on the functions $c$ and $d$ \cite{Z3}. However,
the ADM definitions are strictly valid only for asymptotically flat
space-times, which is not the case of the Bondi-Sachs space-time.

In view of the structure of Eq. (\ref{25}), we may
identify the term that depends only on $c$ and $d$, Eq. (\ref{28}), as the 
gravitational energy of radiation
$E_{rad}$, since the functions $c$ and $d$ are non-vanishing at spacelike 
infinity and the news functions propagate over the whole three-dimensional 
space. The expression of $E_{rad}$ is clearly finite.

By requiring $d=0=\bar{l}$, we restrict the Bondi-Sachs space-time to the Bondi 
space-time, which is endowed with axial symmetry. In this space-time, a
suggested form for the news function $c(u,\theta)$ is given by Eq.
(\ref{34}). The functions $c$ and $d$ yield the news functions, which are
assumed as the radiating degrees of freedom of the gravitational field. From 
this point of view, it is interesting that the energy of gravitational radiation
(\ref{28}) results in a quadratic combination of the coefficients $a_n$. \par
\bigskip

\noindent{\bf Acknowledgement}\par
\noindent The authors are grateful to K. H. C. Castello-Branco for indicating
Ref. \cite{Hobill}.

\end{document}